\documentclass[reqno,a4paper]{amsart}
\usepackage{amsaddr}
\usepackage{enumerate}
\usepackage{graphicx}
\usepackage[utf8]{inputenc}
\usepackage{amsmath,amssymb,MnSymbol,dsfont}
\usepackage{hyperref}
\usepackage{doi}
\usepackage[numbers,sort&compress]{natbib}
\usepackage{multicol}
\usepackage{tgtermes}
\usepackage{mathtools}
\usepackage{color}

\newcommand{\eps}{\varepsilon}
\newcommand{\EE}{\mathbb{E}}
\newcommand{\RR}{\mathds{R}}

\theoremstyle{definition}

\begin{document}

\title{Metadynamics for transition paths in irreversible dynamics}
\author{Tobias Grafke}
\address{University of Warwick, Coventry CV4 7AL, United Kingdom}
\email{T.Grafke@warwick.ac.uk}
\author{Alessandro Laio}
\address{SISSA, Via Bonomea 265, I-34136 Trieste, Italy and ICTP, Strada Costiera 11, 34014 Trieste, Italy}
\email{laio@sissa.it}

\begin{abstract}
  Stochastic systems often exhibit multiple viable metastable states
  that are long-lived. Over very long timescales, fluctuations may
  push the system to transition between them, drastically changing its
  macroscopic configuration. In realistic systems, these transitions
  can happen via multiple physical mechanisms, corresponding to
  multiple distinct transition channels for a pair of states. In this
  paper, we use the fact that the transition path ensemble is
  equivalent to the invariant measure of a gradient flow in pathspace,
  which can be efficiently sampled via metadynamics. We demonstrate
  how this pathspace metadynamics, previously restricted to reversible
  molecular dynamics, is in fact very generally applicable to
  metastable stochastic systems, including irreversible and
  time-dependent ones, and allows to estimate rigorously the relative
  probability of competing transition paths. We showcase this approach
  on the study of a stochastic partial differential equation
  describing magnetic field reversal in the presence of advection.
\end{abstract}

\maketitle
\section{Introduction}

Rare events in stochastic systems lie at the core of many applications
in the sciences, describing phenomena as wide as regime changes and
tipping points in Earth’s
climate~\cite{hoffman-kaufman-halverson-etal:1998,
  alley-marotzke-nordhaus-etal:2003,
  pierrehumbert-abbot-voigt-etal:2011, ashwin-heydt:2020},
conformation changes of biomolecules and protein
folding~\cite{levinthal:1968, bryngelson-onuchic-socci-etal:1995},
material sciences~\cite{e-ren-vanden-eijnden:2003},
genetic switches~\cite{gardner-cantor-collins:2000,
  allen-warren-wolde:2005}, etc. These applications have in common
that the underlying stochastic system spends very long times close to
one macroscopic state, only to eventually switch to a completely
different regime due to small and random fluctuations. From the
perspective of stochastic analysis, these rare but important
transitions correspond to a stochastic bridge, i.e. a realization of
the process that starts in one but ends in another basin of
attraction. A whole class of rare event algorithms have been developed
to obtain information about transition paths, from importance
sampling, over genealogical particle algorithms to Freidlin-Wentzell
theory and large deviations. They are valid in different regimes, and
have different ranges of applicability
(see~\cite{grafke-vanden-eijnden:2019} for an overview).

Due to the utmost importance of rare events in applications, several
methods have been developed to enhance the probability of observing
such events in computer simulations. Transition path sampling (TPS)
~\cite{bolhuis-chandler-dellago-etal:2002, borrero-dellago:2016,
  bolhuis-swenson:2021} and string-based
methods~\cite{henkelman-jonsson:2000, e-ren-vanden-eijnden:2002} allow
focusing the computational effort in the simulation of the transition
events, avoiding spending time in sampling the fluctuations of the
system within the metastable states. Umbrella
sampling~\cite{torrie-valleau:1977, kumar-rosenberg-bouzida-etal:1992,
  kastner:2011}, adaptive force bias~\cite{darve-pohorille:2001,
  darve-wilson-pohorille:2002, henin-chipot:2004,
  zhao-fu-lelievre-etal:2017} and
metadynamics~\cite{laio-parrinello:2002,
  barducci-bussi-parrinello:2008, bussi-laio:2020}, allow
reconstructing the free energy landscapes of metastable systems, which
are characterized by the presence of at least two deep free energy
minima.
 Free energy methods require a preliminary choice of a
small set of collective variables (CVs) which are supposed at least to
distinguish all the metastable states. It can be shown that if there
are precisely two metastable states then the optimal CV is the
committor function~\cite{vanden-eijnden:2006,
  helfmann-riberaborrell-schuette-etal:2020}, and many methods have
recently been developed using machine learning and data science
techniques in order to deduce or approximate
it~\cite{thiede-giannakis-dinner-etal:2019,
  finkel-webber-gerber-etal:2021, lucente-herbert-bouchet:2022,
  lucente-rolland-herbert-etal:2022}.  TPS, if used only to simulate
the transitions, does not require choosing a CV, but some of its
variants aimed at estimating the
rate~\cite{TIS,allen-valeriani-wolde:2009} also require using a
CV. Most TPS-based techniques suffer from metastability in path space:
if multiple competing transition mechanisms exist, the trajectories
generated by TPS typically remain confined in a single reaction
channel. To address this problem it has been suggested to perform
metadynamics in path space~\cite{borrero-dellago:2016,
  mandelli-hirshberg-parrinello:2020, bolhuis-swenson:2021}. The idea,
introduced in these works, is driving the transition paths towards
unexplored reaction channels by a history-dependent bias defined as a
function of a CV capable of distinguishing those channels.

Most of these approaches were developed specifically for applications
in molecular and atomistic simulations. In these systems the dynamics
(typically) satisfies detailed balance, and samples a stationary and
currentless probability measure. However, many stochastic models
describing important physical systems do not satisfy these properties
and regularly detailed balance is violated [cite]. The dynamics does
not necessarily sample a stationary probability measure, and even when
this is the case this probability measure is often not
currentless. Stochastic Partial Differential Equations (SPDEs), which
are used for example for describing the mesoscopic dynamics of fluids,
active matter system, and atmosphere or ocean dynamics, often include
explicitly time-dependent terms. In these condition, defining a free
energy is not possible, and most of the methods for studying rare
events cannot be applied in their standard form.

In this work we propose a principled approach which allows exploring
the bridge ensemble for any kind of dynamics, including irreversible
dynamics. The approach is based on the observation that the path
probability density defined by the Onsager-Machlup-Jacobi (OMJ)
functional defines a currentless probability measure even if the
stochastic dynamics from which the action is derived does not
\cite{Faccioli97,passerone2001action}. By performing Langevin dynamics
in path space one can generate an ensemble of paths with a probability
proportional to the exponential of the OMJ action.  This allows
defining a meaningful free energy as a function of virtually any
collective variable, and use enhanced sampling methods to estimate
this free energy.  In particular, in order to explore the free energy
landscape in path space we use metadynamics, following
ref. \cite{mandelli-hirshberg-parrinello:2020}. This approach allows
studying rare events generated by irreversible and time-dependent
stochastic processes and characterized by multiple competing
transition mechanisms.  In the context of irreversible dynamics, the
advantage of this approach is that we can rely on the rigorously
derived equivalence of the bridge measure and the invariant measure of
certain stochastic partial differential equations, as proposed
in~\cite{stuart-voss-wilberg:2004, hairer-stuart-voss:2007,
  beskos-roberts-stuart-etal:2008} and formally generalized to
arbitrary non-reversible systems~\cite{hairer-stuart-voss:2007}.

In the following, we will first introduce into bridge path sampling in
section~\ref{sec:bridge-path-sampling}, and subsequently describe the
metadynamics algorithm generally in section~\ref{sec:metadynamics}. We
then show how to apply metadynamics to transition paths to sample
multiple competing transition mechanisms in
section~\ref{sec:pathsp-metadyn}. Section~\ref{sec:numerical-examples}
then demonstrates the applicability of the algorithm to a multitude of
examples, from a temporally oscillating doublewell in
section~\ref{sec:two-comp-trans}, to a two-dimensional irreversible
stochastic differential equation (SDE) in
section~\ref{sec:oscill-doubl}, and finally in
section~\ref{sec:magn-field-revers} the stochastic Ginzburg-Landau
equation with background advection as a field-theoretic example with
spatial extent. We conclude with a discussion in
section~\ref{sec:discussion}.

\section{Metadynamics in Pathspace}
\label{sec:metadyn-pathsp}

\subsection{Bridge Path Sampling and Transition Path Sampling}
\label{sec:bridge-path-sampling}

Consider for $X_t\in\RR^n$ the stochastic differential equation
\begin{equation}
  \label{eq:SDE}
  dX_t = b(X_t,t)\,dt + \sqrt{2\eps}\,dW_t\,,\quad t\in[0,T]\,,\quad X_0=x_-\,,
\end{equation}
where $b:\RR^n\times[0,T]\to\RR^n$ is the deterministic drift vector
field, and $W_t$ is $\RR^n$-dimensional Brownian motion. We are
interested in drawing sample paths from~(\ref{eq:SDE}), but
conditional on a (possibly rare) \emph{final condition}
$X_T=x_+$. Such trajectories are called \emph{stochastic bridges}, as
they connect $x_-$ to $x_+$ via the stochastic
process~(\ref{eq:SDE}). Sampling such bridges, i.e.~drawing paths from
the \emph{bridge ensemble}, is non-trivial, as simply integrating
forward-in-time the stochastic equation~(\ref{eq:SDE}) has a very low
probability of arriving anywhere near $x_+$, in particular in high
dimension and/or if the final point is difficult to reach for the
dynamics (i.e.~if hitting near $x_+$ is \emph{rare}).

Traditional examples for bridge sampling problems include molecular
dynamics and chemical reactions, where we are interested in
transitions between a reactant and product state of chemical
molecules. Here, a well-developed theory of the form of transition
state theory (TST) and transition path sampling (TPS)
exists. Many methods employed in TPS, though, assume
\emph{reversibility} of the underlying stochastic process (with some
notable exceptions, such as forward flux
sampling~\cite{allen-valeriani-wolde:2009}), which is indeed fulfilled
in the classical molecular dynamics case. For equation~(\ref{eq:SDE}),
reversibility corresponds to demanding that the drift is of the
particular form $b(x) = -\nabla U(x)$ for a potential
$U:\RR^n\to\RR$. In that case, the stationary density is explicitly
known to be the Gibbs
distribution~$\rho(x)\propto\exp(-\eps^{-1}U(x))$.

For general drifts $b$, i.e.~for the \emph{irreversible} setting,
bridge path sampling can be attempted as well, for example in the form
of pathspace Langevin Markov Chain Monte Carlo
(MCMC)~\cite{stuart-voss-wilberg:2004, hairer-stuart-voss:2007,
  beskos-roberts-stuart-etal:2008}, which at least formally allows a
generalization to non-reversible
systems~\cite{hairer-stuart-voss:2007}. This is the path we will take
in the following.

\subsubsection{The Onsager-Machlup functional and large deviation theory}
\label{sec:onsag-machl-jacobi}

In the vanishing noise limit, $\eps\to0$, transitions from $x_-$ to
$x_+$ of~(\ref{eq:SDE}) are described by the minimizers, also termed
\emph{instantons}, of the Freidlin-Wentzell action
functional~\cite{freidlin-wentzell:2012},
\begin{equation}
  \label{eq:FW}
  I[\phi] = \tfrac12\int_0^T |\dot\phi - b(\phi)|^2\,dt\,.
\end{equation}
This is because the probability of observing any given path $\phi(t)$
is given by $P(\phi(t))\sim\exp(-\eps^{-1} I[\phi])$, as can be
intuited by ``solving'' the SDE~(\ref{eq:SDE}) for the noise, which
has a Gaussian distribution. Thus, considering all possible transition
pathways $\phi(t)$ between $x_+$ and $x_-$, the minimizer will
exponentially dominate all other scenarios. In other words, for
vanishing noise, the bulk of the transition trajectories between $x_-$
and $x_+$ will lie in the vicinity of the instanton trajectory. These
minimizers can be computed efficiently by numerical
means~\cite{e-ren-vanden-eijnden:2004, heymann-vanden-eijnden:2008,
  grafke-grauer-schaefer-etal:2014, grafke-grauer-schaefer:2015,
  grafke-grauer-schindel:2015, grafke-vanden-eijnden:2019}.

For finite noise, this functional must be amended by an
additional term~\cite{pinski-stuart:2010} to obtain the
Onsager-Machlup functional
\begin{equation}
  \label{eq:OM}
  S[\phi] = \tfrac12\int_0^T\left(|\dot\phi - b(\phi)|^2 + \eps \nabla\cdot b(\phi)\right)\,dt\,.
\end{equation}
While finite noise minimizers of the Onsager-Machlup
function~(\ref{eq:OM}) can be found, for example by performing a
gradient descent in trajectory space, a key realization is that,
similar to above, paths $\phi(t)$ of the process~(\ref{eq:SDE}) for
finite noise can formally be thought of as drawn from a Gibbs-type
distribution
\begin{equation}
  P[\phi(t)]\sim \exp(-\eps^{-1} S[\phi])\,.
\end{equation}
As a consequence, one can sample the bridge ensemble by instead
integrating the stochastic \emph{partial} differential equation
\begin{equation}
  \label{eq:SPDE-abstract}
  \partial_\tau \phi = -\frac{\delta S[\phi]}{\delta \phi} + \sqrt{2\eps}\eta\,,
\end{equation}
where $\phi(t,\tau) : [0,T]\times[0,\infty] \to \RR^n$ is a field in
physical time $t$ and \emph{virtual time} $\tau$, with boundary
conditions $\phi(0,\tau)=x_-$, $\phi(T,\tau)=x_+$. Here, $\eta(t,\tau)$
is white in virtual and physical time, i.e.
\begin{equation}
  \EE \eta(t,\tau)\eta(t',\tau') = \delta(\tau-\tau')\delta(t-t')\,.
\end{equation}
Effectively, the physical time $t$ is treated as spatial variable, and
the evolution is happening in the newly introduced virtual time
variable. Sampling the SPDE~(\ref{eq:SPDE-abstract}) amounts to
Langevin Markov chain Monte Carlo (MCMC) in path space, i.e.~the SPDE
corresponds to a Markov process~(namely the functional Langevin
equation) with an invariant measure equivalent to the original
transition path ensemble.

If the SDE~(\ref{eq:SDE}) is in one dimension, $n=1$, then
combining~(\ref{eq:OM}) and~(\ref{eq:SPDE-abstract}) yields the SPDE
\begin{equation}
  \partial_\tau \phi = \partial_t^2 \phi - b(\phi)b'(\phi) - \tfrac12\eps b''(\phi) + \sqrt{2\eps}\eta\,,\quad \phi(0,\tau)=x_-\,,\quad \phi(T,\tau)=x_+\,,
\end{equation}
which is essentially a gradient flow with noise, with the gradient
computed via the Euler-Lagrange equation of~(\ref{eq:OM}). This result
has been rigorously derived in~\cite{stuart-voss-wilberg:2004,
  hairer-stuart-voss:2007, beskos-roberts-stuart-etal:2008}.

If $n>1$ and the system is non-gradient, the Euler-Lagrange equation
of the the action~(\ref{eq:OM}) instead yields
\begin{equation}
  \label{eq:SPDE-full}
  \partial_\tau\phi = \partial_t^2\phi - (\nabla b(\phi) - \nabla b(\phi)^\dagger)\partial_t\phi - b(\phi)\nabla b(\phi) - \tfrac12\eps\nabla (\nabla\cdot b(\phi)) + \sqrt{2\eps}\eta\,,
\end{equation}
subject to
\begin{equation}
  \phi(0,\tau)=x_-\,,\quad \phi(T,\tau)=x_+\,,
\end{equation}
where the additional term proportional to $\partial_t\phi$ vanishes in
the case of detailed balance, where $\nabla b(\phi)$ is
self-adjoint. This SPDE has been conjectured
in~\cite[sec.~9]{hairer-stuart-voss:2007}, and also used
in~\cite{drummond:2017}.

Indeed,~(\ref{eq:SPDE-abstract}) can be used to effectively sample the
bridge ensemble, since every generated path automatically fulfills the
initial and final conditions of the bridge and further the SPDE has
the bridge ensemble as invariant measure. The problem that motivates
this work is that the Onsager-Machlup action~(\ref{eq:OM}) itself
might have multiple local minima in trajectory space, each
corresponding to a locally effective transition trajectory. This is
the case, for example, when there are multiple competing reaction
channels, or multiple physical mechanisms for the transition to
occur. In that case, bridge sampling via~(\ref{eq:SPDE-abstract}) will
be inefficient, since the SPDE will spend exponentially long times in
one of the minima, and only very rarely explore others. We here
propose to overcome this limitation by introducing metadynamics to the
sampling of~(\ref{eq:SPDE-abstract}).

\subsection{Metadynamics}
\label{sec:metadynamics}

Metadynamics is a rare event simulation technique that allows
effective estimation of the free energy even in the presence of
multiple local minima with large barriers. It was originally developed,
and is mostly used, to sample high-dimensional atomistic systems as
present in molecular dynamics, with applications in e.g.~material
science or biophysics.

Its main idea is to avoid a Markov Chain Monte Carlo sampler to get
stuck in a single maximum of the Gibbs measure, by gradually 'filling'
the current free energy minimum until saturation, so that the system
eventually starts exploring all other states. This filling is done in
a controlled way that later allows to reconstruct the original
probability landscape. A multitude of variants of metadynamics
exist~\cite{bussi-laio:2020}, but they are restricted in application
to reversible systems, i.e.~those with an underlying free energy
landscape that can be filled.

The key realization of this paper is the fact that even for
irreversible systems of the form~(\ref{eq:SDE}), the corresponding
transition path sampling problem is a gradient flow \emph{in path
  space} and thus amenable to metadynamics. In short, while the
dynamics might not have an underlying potential landscape, its
\emph{trajectories} do, in the form of the Onsager-Machlup
action~(\ref{eq:OM}).

\subsubsection{Well-tempered metadynamics}
\label{sec:well-temp-metadyn}

In order to implement metadynamics, consider a reversible stochastic
process $q_t\in\RR^n$ evolving according to
\begin{equation}
  \label{eq:gradient-dynamics}
  dq_t = -\nabla U(q_t)\,dt + \sqrt{2\eps}\,dW_t\,,
\end{equation}
implying that the stationary distribution of $q_t$ is given by the
Gibbs distribution $q\sim\rho(q) \propto \exp(-\eps^{-1} U(q))$. In
metadynamics, the gradient dynamics~(\ref{eq:gradient-dynamics}) are
augmented by a \emph{biasing potential} $V(q,t):\RR^n\times[0,T]\to
\RR$ via
\begin{equation}
  \label{eq:gradient-dynamics-with-bias}
  d q_t = -\nabla U(q_t)\,dt - \nabla V(q_t,t) + \sqrt{2\eps}\,dW_t\,,
\end{equation}
where the biasing potential is large (and thus repelling) in regions
where the system spent a lot of time in the past. That way,
exploration is increasingly encouraged over time.

The time-dependent bias potential is assumed to depend on the
coordinates $q$ only via a smooth function $f:\RR^n \to \RR^m$ with
$m\ll n$ (in typical applications $m<4$). The function $f$ should be
chosen in such a way that the marginal distribution with respect to
$f$,
\begin{equation}
  \rho_f(s) = \int_{\RR^n} \rho(x)\delta(s-f(x))\,dx
\end{equation}
is \emph{multimodal}, namely characterized by the presence of at least
two distinct maxima, separated by a region in which $\rho_f(s)$ is
small. The function $f$ is typically called Collective Variable (CV),
as it is assumed to be able to recapitulate the most salient features
of the probability density.

A typical protocol for building up the biasing potential is
\emph{well-tempered} metadynamics, in which the biasing potential and
process co-evolve according to
\begin{equation}
  \label{eq:well-tempered-evolution-CV}
  \begin{cases}
    dq_t &= -\nabla U(q_t)\,dt - \nabla (V\circ f)(q_t,t)\,dt + \sqrt{2\eps}\,dW_t\,,\\
    \dot V(s,t) &= w e^{-V(s,t)/\kappa} \psi_\delta(f(q_t)-s)\,,\qquad V(s,0) = 0\,.
  \end{cases}
\end{equation}
The above intuitively deposits droplets of shape $\psi_\delta$
(typically Gaussian profiles of width $\delta$) at the current
position of the CV, $f(q_t)$, with a weight $w>0$. The weight is
exponentially decreasing for increasing $V(s,t)$, with a scale given
by $\kappa>0$ 
(see~\cite{barducci-bussi-parrinello:2008, bussi-laio:2020}).
In the long-time limit, the biasing potential $V_\infty(s)$  yields an estimate 
of $\rho_f(s)$ via
\begin{equation}
  \label{eq:V-infty}
  \rho_f(s) = e^{\frac{\kappa+\eps}\kappa V_\infty(s)}\,.
\end{equation}
As a consequence, $\kappa$ can be
seen as an effective \emph{sampling temperature} or noise amplitude,
different from the physical noise amplitude given by $\eps$,
and~(\ref{eq:V-infty}) translates between the two.

\subsection{Pathspace Metadynamics}
\label{sec:pathsp-metadyn}

Applying now well-tempered
metadynamics~(\ref{eq:well-tempered-evolution-CV}) in collective
variables to the Onsager-Machlup stochastic gradient
flow~(\ref{eq:SPDE-abstract}) yields an effective method for sampling
transitions in the presence of multiple possible transition
channels. This technique was presented
in~\cite{mandelli-hirshberg-parrinello:2020} in the context of
equilibrium systems from molecular dynamics. We will derive this
technique in our general notation in order to sample transition paths
in nonequilibrium systems.

Consider a CV function $f[\phi]: C^1(\RR^n,[0,T])\to\RR^m$,
reducing from continuous trajectories on $\RR^n\times[0,T]$ to $m$
collective variables. For example, one could consider the location of
the trajectory at $t=T/2$, which corresponds to $f[\phi] = \phi(T/2)$
and thus $\delta f/\delta \phi = \delta(t-T/2)$.

Then, for a trajectory $\phi(t)$, integrate, in
virtual time $\tau$,
\begin{equation}
  \label{eq:pathspace-metadynamics}
  \begin{cases}
    \partial_\tau \phi(t,\tau) &= -\frac{\delta S[\phi]}{\delta \phi} - \frac{\delta V\circ f([\phi],\tau)}{\delta \phi} + \sqrt{2\eps} \eta(t,\tau)\,,\\
    \dot V(s,\tau) &= w e^{-V(s,\tau)/\kappa} \psi_\delta(f[\phi(t,\tau)]-s)\,.
  \end{cases}
\end{equation}
We term the above \emph{pathspace metadynamics}, and
will use it in the following in several examples of transitions
in irreversible processes that exhibit multiple transition channels.

\section{Numerical Examples}
\label{sec:numerical-examples}

\subsection{Two competing transition paths}
\label{sec:two-comp-trans}

\begin{figure}
  \begin{center}
    \includegraphics[width=0.9\textwidth]{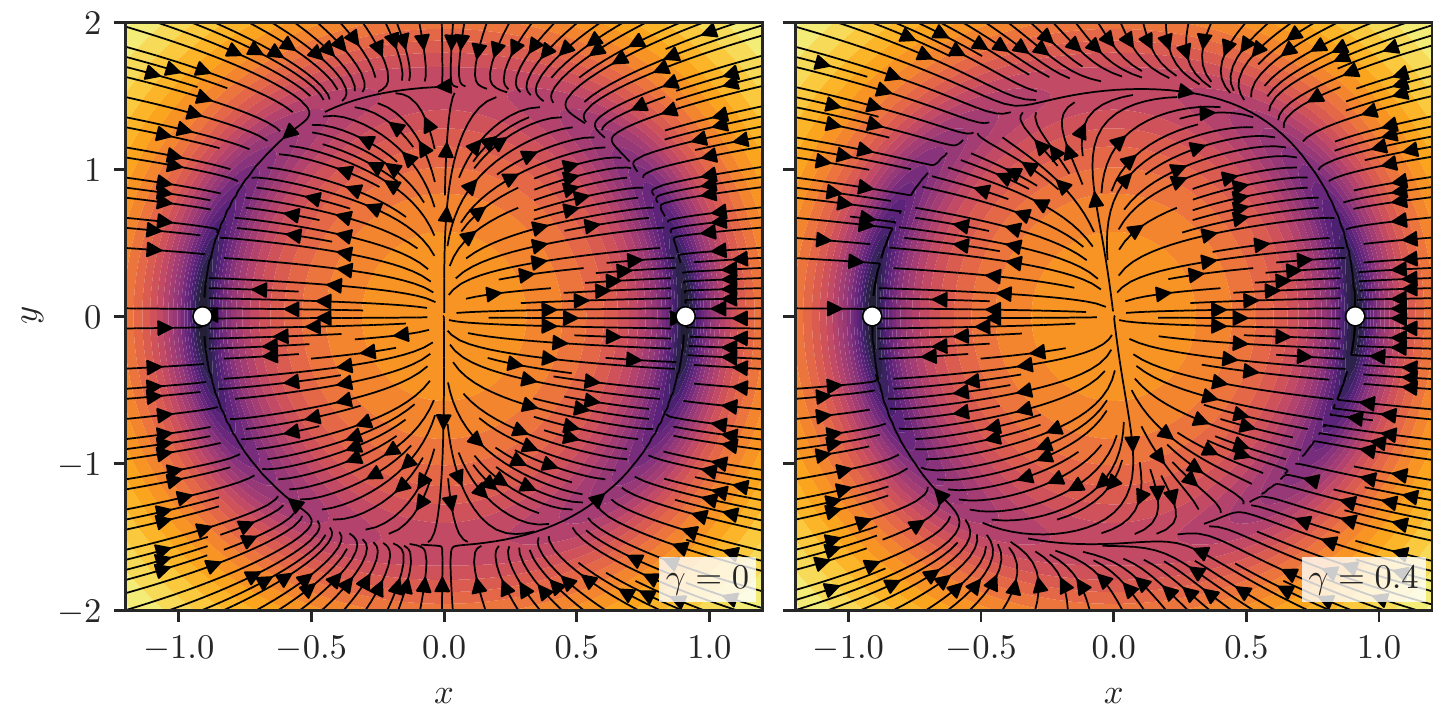}
  \end{center}
  \caption{The 2-dimensional test model, as defined by
    equation~(\ref{eq:2d-model}). For $\gamma=0$ (left), the drift is
    a gradient flow in the potential~(\ref{eq:2d-potential}). For
    nonzero $\gamma$, such as $\gamma=0.4$ (right), an additional
    swirl is added. There are two fixed points of the system (white
    dots), and notably \emph{two distinct} most likely transition
    pathways between them (upper and lower). Non-vanishing swirl
    makes one transition path preferred over the other. In general it
    is hard to know the relative likelihood of the two transition
    channels. }
  \label{fig:2d-model}
\end{figure}

\begin{figure}
  \begin{center}
    \includegraphics[width=\textwidth]{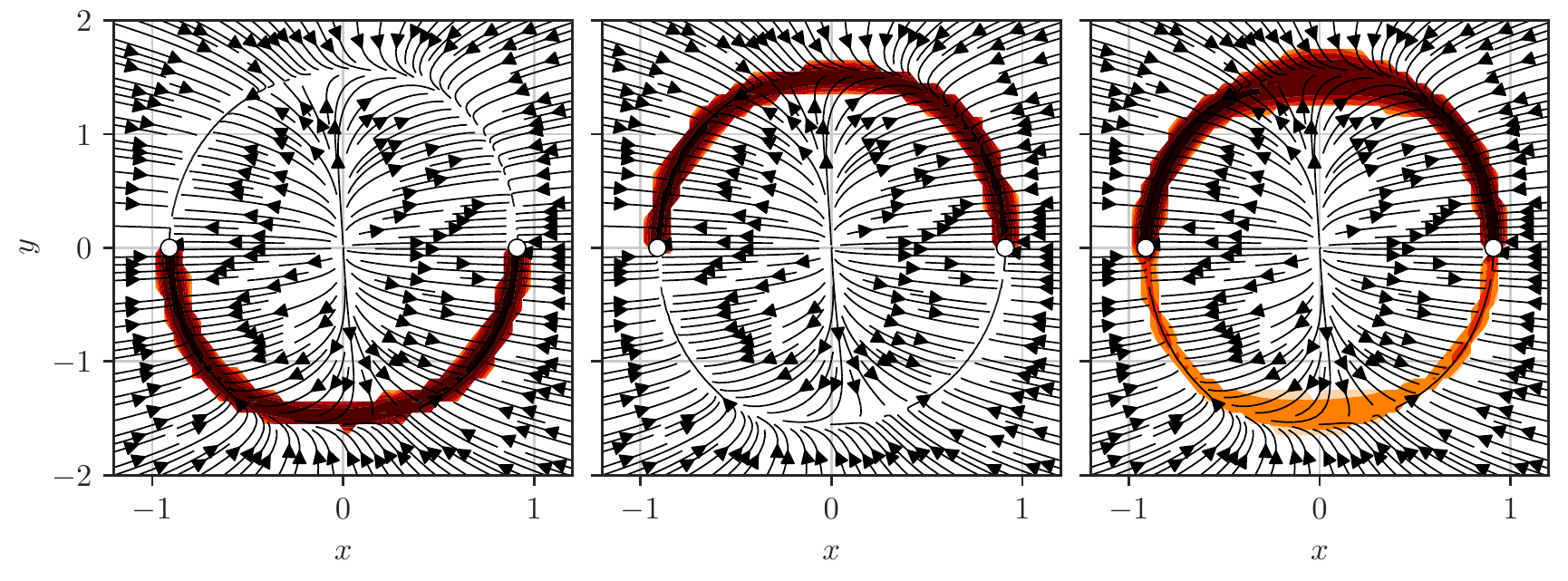}
  \end{center}
  \caption{Trajectories sampled by transition path sampling
    (\emph{left} and \emph{center}) versus trajectories explored by
    pathspace metadynamics (\emph{right}), for the slightly asymmetric
    2d example ($\gamma=0.2$). Traditional transition path sampling
    can only explore one of the two available transition mechanisms,
    depending on the initial transition trajectory fed to the
    algorithm. Metadynamics can explore the whole space of possible
    transitions.\label{fig:2d-mcmc-vs-metadyn}}
\end{figure}

\begin{figure}
  \begin{center}
    \includegraphics[width=\textwidth]{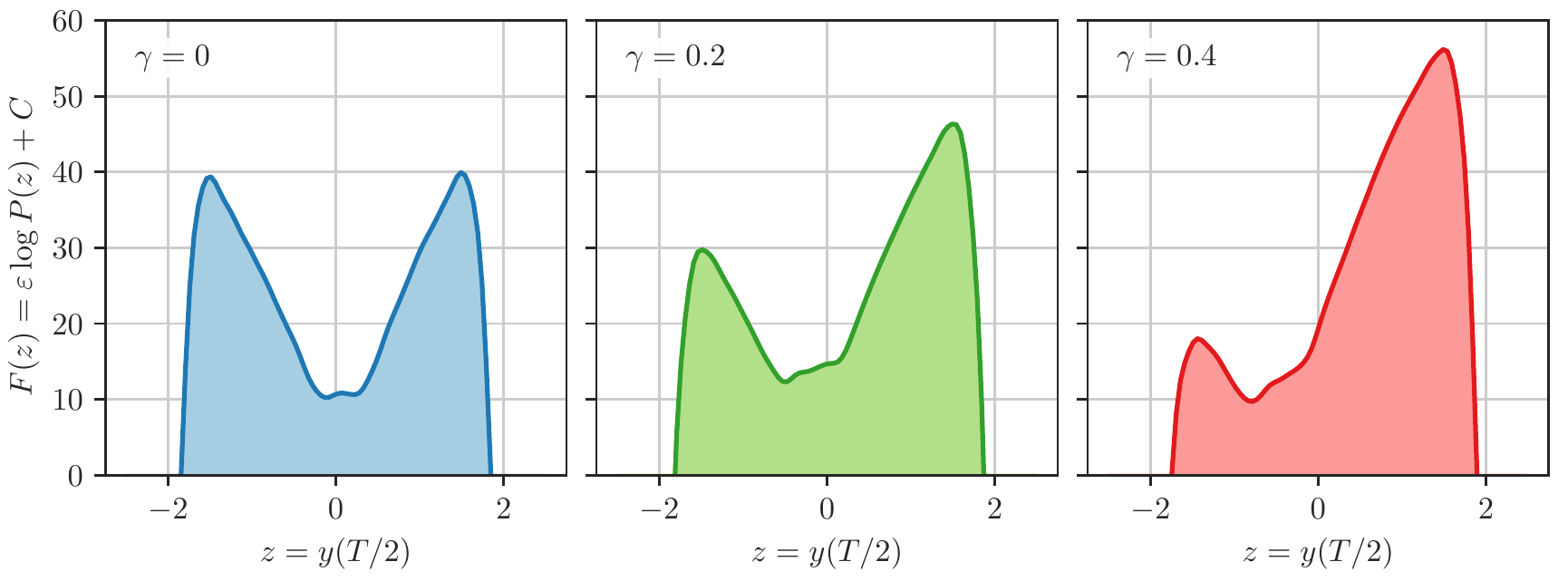}
  \end{center}
  \caption{Probability of taking the upper or lower transition path as
    a function of $\gamma$. Transitioning from left to right, the
    upper transition path (i.e.~$z>0$) is preferred for positive $\gamma$:
    The probability to observe a positive $y$-value at the center of
    the time interval, $y(T/2)>0$, is increasing, but both upper and
    lower channel remain present. Shown is the scaled logarithm of the
    probability, $\eps \log P(z) + C$, with arbitrary normalization
    constant $C$.\label{fig:2d-UBias-for-gammas}}
\end{figure}

The simplest case of coexistence of multiple transition paths is
2-dimensional gradient diffusion in a landscape that has two separate
valleys connecting the local minima. To this end, consider for
$Z=(X,Y)\in\RR^2$ the system
\begin{equation}
  \label{eq:2d-model}
  dZ = -\nabla V(Z)\,dt + l(Z)\,dt + \sqrt{\eps}\,dW
\end{equation}
with
\begin{equation}
  \label{eq:2d-potential}
  V(X,Y) = \tfrac14(X^2-1)^2 + \tfrac14(Y^2 - \alpha(c^2-X^2))^2\,,
\end{equation}
and we choose
\begin{equation}
  \label{eq:2d-additional-drift}
  l(x,y) = \left(
  \begin{matrix}
    \gamma y\\
    0
  \end{matrix}
  \right)\,.
\end{equation}
This system is not reversible for $\gamma\ne0$ and corresponds to a
diffusion in a potential with two minima, $(X_\pm,Y_\pm) =
(\pm\sqrt{(1+\alpha^2c^2)/(1+\alpha^2)}, 0)$, but with an additional
non-gradient force $l(z)$ of strength $\gamma$ added. The potential,
and the flowlines of the full drift, are depicted in
figure~\ref{fig:2d-model}. Notably, there are \emph{two distinct} most
likely transition pathways between the left and the right fixed point,
through the upper or the one lower channel. If $\gamma=0$, both of
these are equally likely due to symmetry. If $\gamma>0$, as visible in
figure~\ref{fig:2d-model} (right), the upper pathway is preferred for
a left-to-right transition. For a general system of this type, it is
hard to compute all possible transition pathways and their relative
likelihood.

Applying pathspace metadynamics to this example corresponds to
integrating equation~(\ref{eq:pathspace-metadynamics}) in virtual time
for trajectories $\phi(t,\tau) = (\phi_x(t,\tau), \phi_y(t,\tau))$
connecting the left to the right fixed point. At every instance in
virtual time, the solution to the SPDE yields a transition path, and
integrating for long times samples from the transition path
ensemble. Without employing metadynamics, the SPDE will quickly
converge to one of the two transition scenarios, and remain stuck
there for exponentially long (virtual) times. Employing metadynamics,
instead, allows us to gradually fill up one of the transition channels
and eventually explore the other as well. In particular in the
scenario $\gamma\ne0$, where the relative likelihood of the two
channels is not obvious, this strategy helps identify their relative
weight.

As collective variable, we choose $m=1$ and $f[\phi] = \phi_y(T/2)$,
i.e.~the $y$-value of the trajectory at the center of the temporal
interval. We expect this value to be positive if the upper channel is
chosen, and negative for the lower. Note though that $f[\phi]=0$ not
necessarily corresponds to a trajectory crossing the domain at the
center. Instead, it might correspond to a trajectory that waits at the
initial point until $t=T/2$ and only afterwards initiates a transition
through either channel. 

Figure~\ref{fig:2d-mcmc-vs-metadyn} shows a comparison of naive
pathspace MCMC (left, center) to pathspace metadynamics (right) for
$\gamma=0.2$. While MCMC results in the exploration of only one of the
two transition mechanisms (either the upper transition or the lower
transition, depending on initialization of the sampler), metadynamics
can efficiently explore both these mechanisms, and compare their
relative likelihood. As expected for positive $\gamma$ the lower
pathway is less likely for a left-to-right transition than the upper
pathway. This picture is confirmed when looking at the probability
distribution of the CV, for different values of $\gamma$, in
figure~\ref{fig:2d-UBias-for-gammas}. While for $\gamma=0$, both upper
and lower pathway are of equal likelihood, for increasing values of
$\gamma$ the upper pathway becomes more likely. Note that the relative
likelihood of the two, for the noise of $\eps=10^{-2}$, makes the
upper pathway exponentially more likely even for rather small positive
$\gamma$. Figure~\ref{fig:2d-UBias-for-gammas} shows the rescaled
log-likelihood.

For this example, we chose $\eps=10^{-2}$, $\kappa=20$, $T=4$,
$\alpha=3$, and $\gamma$ as indicated, with $N_t=64$ discretization
points and $w=1$, and $\psi_\delta(s) = \exp(-s^2/2\delta^2)$ with
$\delta=0.1$. Discretization in time is done via second order finite
differences. Integration in virtual time is done with Euler-Maruyama
with stepsize $\Delta \tau=10^{-3}$.

\subsection{Oscillating doublewell}
\label{sec:oscill-doubl}

\begin{figure}
  \includegraphics[width=0.48\textwidth]{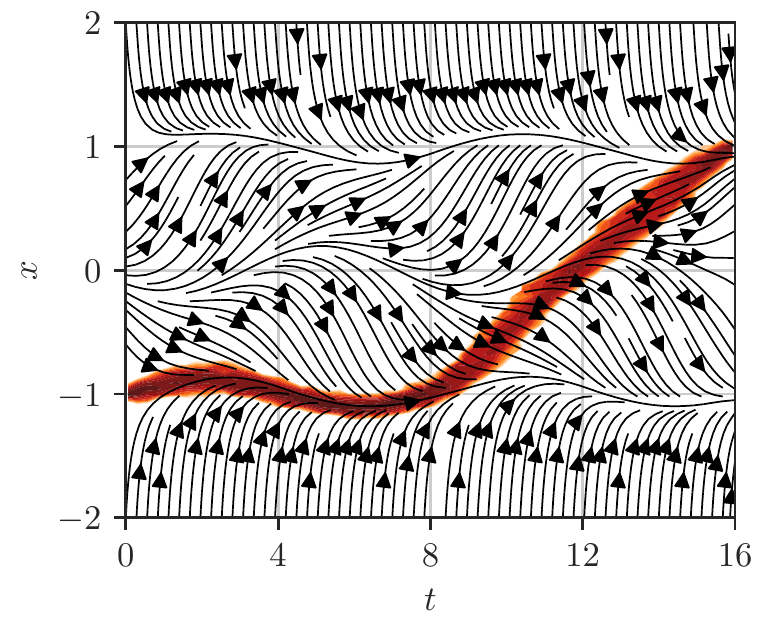}
  \includegraphics[width=0.48\textwidth]{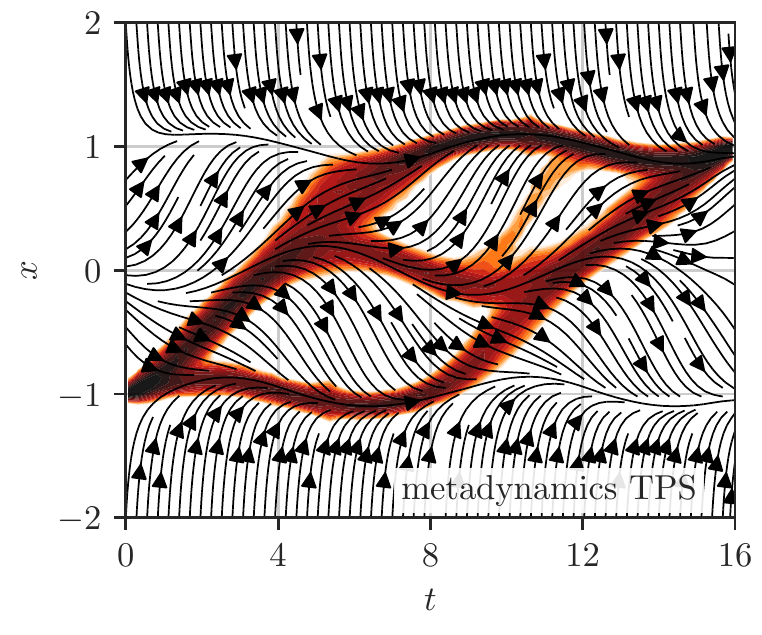}
  \caption{\emph{Left:} Path density plot obtained from pathspace MCMC
    sampling for transitions in the oscillating double well potential
    from $x\!=\!-1$ to $x\!=\!1$, shown in the $x$-$t$-plane. Even for
    a very large number of samples, the trajectory never leaves the
    current local minimum determined by the initial condition of the
    sampling procedure. \emph{Right:} Path density plot obtained from
    pathspace metadynamics for the same system. As clearly visible,
    the process explores all relevant transition trajectories,
    irrespective of the initial condition of initialization of the
    sampling procedure. For both plots, $\eps=10^{-2}$ and $\Delta
    T=2$. \label{fig:moving-doublewell}}
\end{figure}

\begin{figure}
  \begin{center}
    \includegraphics[width=0.48\textwidth]{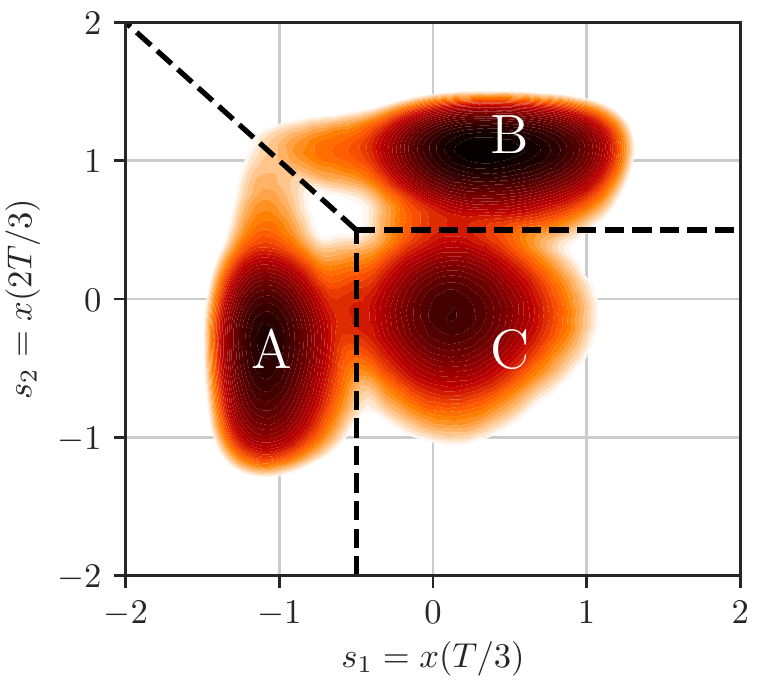}
    \includegraphics[width=0.48\textwidth]{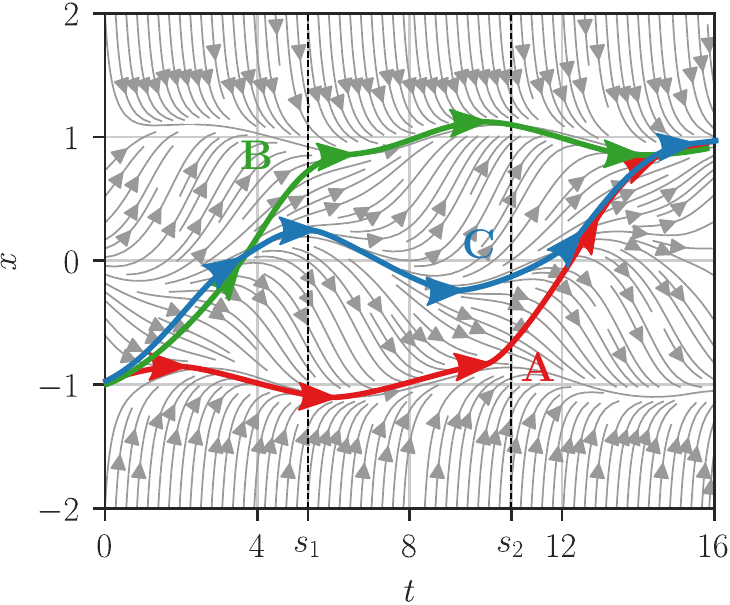}
  \end{center}
  \caption{\emph{Left:} Limiting biasing potential $V_\infty(s)$ in
    the space of collective coordinates $(s_1,s_2) = (x(T/3),x(2T/3))$
    for the moving double well experiment. There are three dominant
    states, $A, B$, and $C$, which correspond to late transition,
    early transition, and dwelling around the saddle,
    respectively. \emph{Right:} Sketch of these three transition
    scenarios. \label{fig:moving-doublewell-Vmeta-2CV}}
\end{figure}

Even in one dimension there can be multiple competing reaction
channels, for example when the system has explicit time dependence and
there are multiple points in time at which a transition is
advantageous. As an example, consider the SDE given in
equation~(\ref{eq:SDE}) with drift
\begin{equation}
  b(x,t) = -x^3 + x + \tfrac14 \sin(4\pi t/T)\,.
\end{equation}
This is simply a diffusion in a double-well potential that is
periodically tilted left or right. In fact, within the time period $T$
the potential tilts exactly twice, allowing for two ``optimal''
opportunities to jump from the left to the right basin. Therefore, the
ensemble of transition paths contains at least two different distinct
trajectories, and sampling both of them efficiently is a hard task.

Concretely, we want to sample trajectories that transition within the
interval $[0,T]$ from the left basin, $X_-=-1$, to the right basin,
$X_+=1$. We could consider employing classical transition path
sampling by integrating $\phi(t,\tau)$ via
equation~(\ref{eq:SPDE-full}) with $\phi(0,\tau)=X_-$ and
$\phi(T,\tau)=X_+$. Then, depending on the initial choice of
transition path, we would sample almost certainly only one of the
multiple possible transition pathways. This is depicted in
figure~\ref{fig:moving-doublewell} (left), which shows as streamlines
the deterministic, time-dependent drift $b(x,t)$, and in color a
weighted histogram of observed samples. If instead we perform
transition path sampling with the help of pathspace metadynamics then
we see how the whole space of feasible transition paths is
explored. In particular, it becomes clear that there are at least
three noteworthy classes of transition trajectories: those that
transition at the earliest opportunity, those at the latest, and those
that spend one oscillation in close proximity to the saddle point
(which, despite its instability, turns out to be a very relevant
transition trajectory). This is shown in
figure~\ref{fig:moving-doublewell} (right), where again we show a
histogram of transition paths $\{x(t)\}_{t=0}^T$.

As CVs, we choose
\begin{equation}
  f[\phi] = (s_1, s_2) = (\phi(T/3), \phi(2T/3))\,,
\end{equation}
i.e.~the location of the sampled trajectory at $1/3$ and $2/3$ of the
time interval. We obtain, after running the
algorithm~(\ref{eq:pathspace-metadynamics}) sufficiently long, a clear
separation of the available options, as visible in the converged
biasing potential $V_\infty(s)$ in
figure~\ref{fig:moving-doublewell-Vmeta-2CV} (left): Either $s_1$
remains small, corresponding to a late transition and remaining in the
original basin for the first oscillation, labeled as region
$A$. Alternatively, $s_2$ is large, corresponding to an early
transition where the trajectory is in the second basin already during
the second oscillation, labeled as region $B$. Lastly, both $s_1$ and
$s_2$ are of moderate value, corresponding to an early transition to
the saddle, but then dwelling at the saddle for a full oscillation,
labeled as region $C$. The biasing potential gives an estimate of the
relative likelihood of these options. These alternatives are sketched
in figure~\ref{fig:moving-doublewell-Vmeta-2CV} (right).

For this example, we chose $\eps=10^{-2}$,
$\kappa=2$, $T=16$, with $N_t=128$ discretization points and $w=1$,
$\delta=0.05$. Discretization in time is done via second order finite
differences. Integration in virtual time is done with Euler-Maruyama
with stepsize $\Delta \tau=10^{-3}$.

\subsection{Magnetic field reversal in the presence of advection}
\label{sec:magn-field-revers}

\begin{figure}
  \begin{center}
    \includegraphics[width=\textwidth]{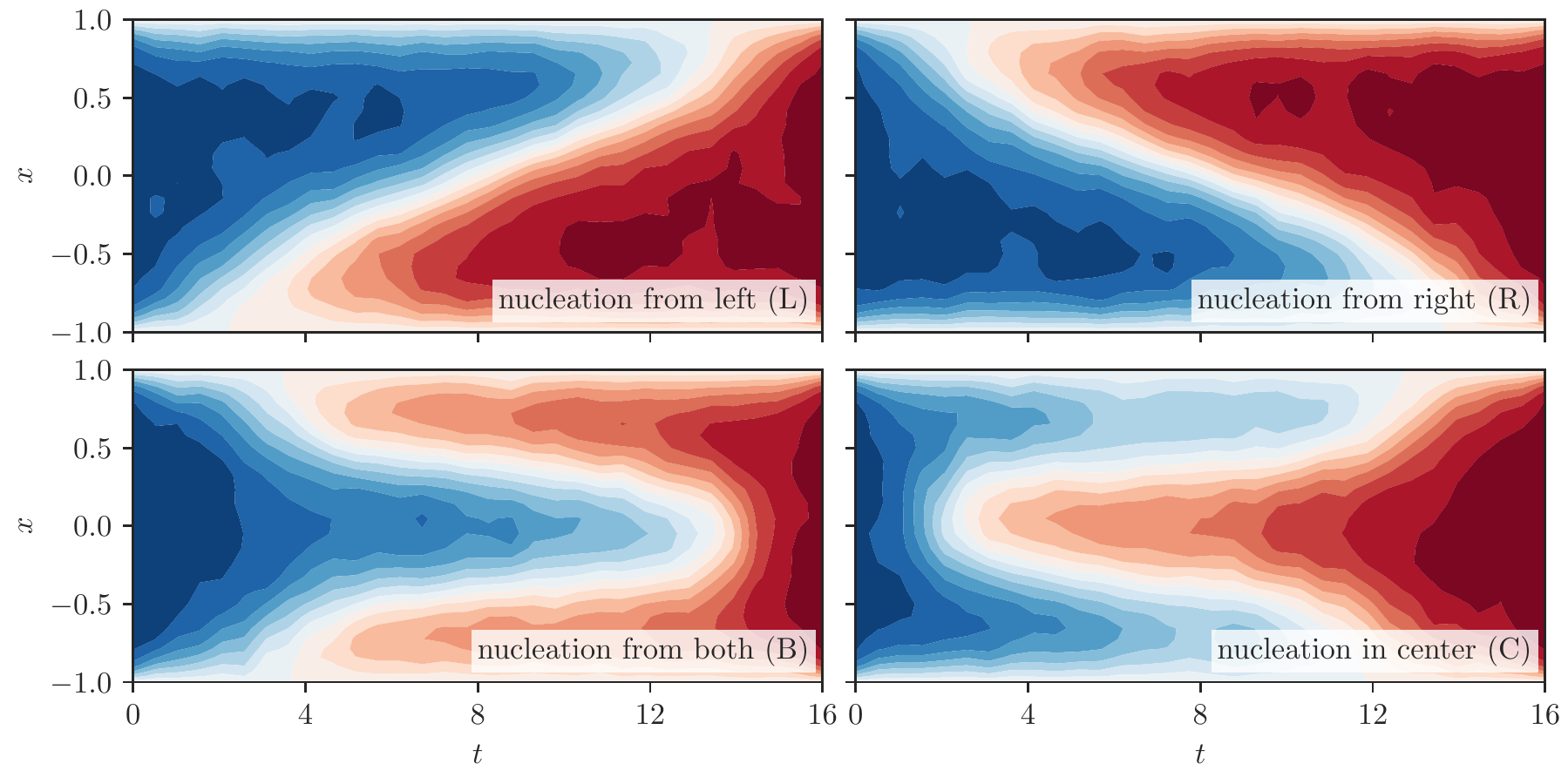}
  \end{center}
  \caption{Snapshots of samples of the four possible transition
    pathways from the negative configuration $u_-(x)$ to the positive
    configuration $u_+(x)$: nucleation from left wall (L), nucleation
    from right wall (R), nucleation from both walls (B), or nucleation
    in the center (C).\label{fig:ginzburg-landau-gamma-0}}
\end{figure}

\begin{figure}
  \begin{center}
    \includegraphics[width=0.48\textwidth]{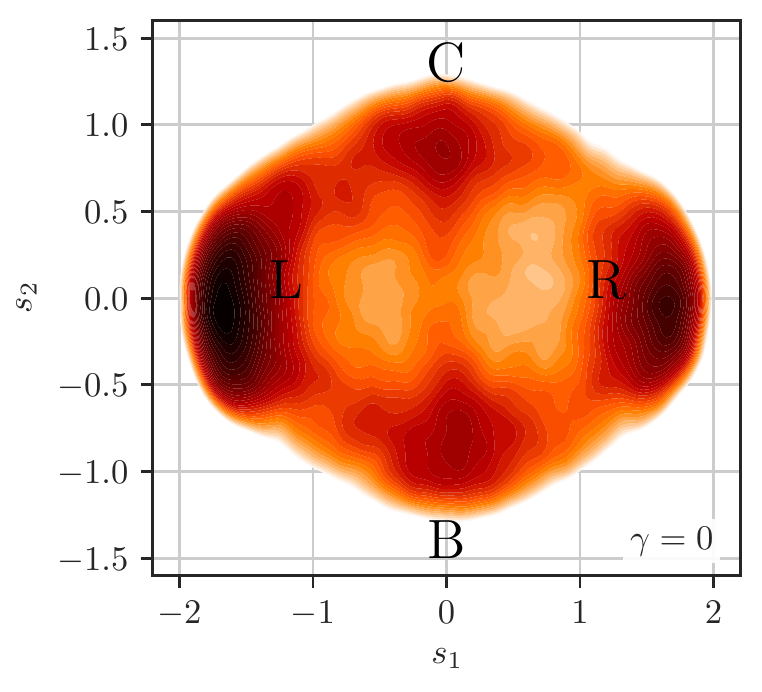}
    \includegraphics[width=0.48\textwidth]{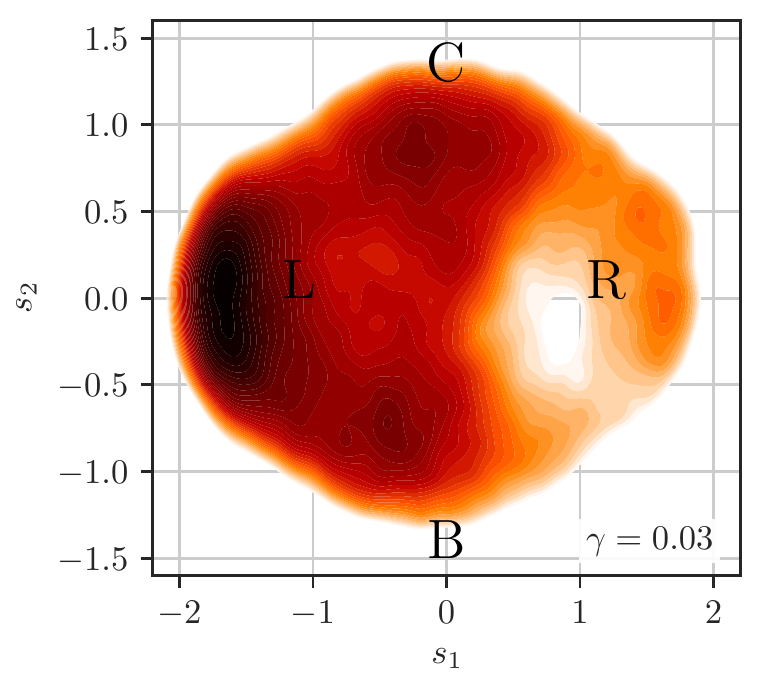}
  \end{center}
  \caption{Left: Limiting biasing potential $V_\infty(s)$ in CV space
    $(s_1, s_2)$, corresponding to the $\sin$ and $\cos$-components at
    $t=T/2$. Intuitively, $s_1$ corresponds to the left-right
    asymmetry, and $s_2$ to the center-boundary asymmetry of the
    configuration in space. Four separate transition channels are
    identified (R, L, C, B). Right: The same, but in the presence of
    background flow, $\gamma=0.03$, as in
    equation~(\ref{eq:ginzburg-landau-shear}). Since fronts are
    propagated predominantly left to right, the nucleation from the
    left boundary (L) is now clearly preferred over nucleation from the
    right (R).
    \label{fig:ginzburg-landau-UBias}}
\end{figure}

To demonstrate the feasibility of applying pathspace metadynamics to
systems with a large number of variables, we will lastly consider a
system with infinitely many degrees of freedom, a stochastic partial
differential equation. Concretely, consider the Ginzburg-Landau (or
equivalently $\phi^4$- or Allen-Cahn) equation for a field
$u(x,t):[-L,L]\times[0,T]\to\RR$ given by
\begin{equation}
  \partial_t u(x,t) = \nu\partial_x^2 u - u^3 + u + \sqrt{2\eps} \eta(x,t)\,,\quad u(-L,t)=u(L,t)=0\,.
\end{equation}
Here, at each point $x$ in space, $u(x,t)$ is traveling in a
double-well potential, driving locally towards $\pm 1$. The diffusion
constant $\nu$ couples neighboring locations, and $\eta(x,t)$ is white
in space and time stochastic noise. The system prefers to be in one of
the two states $u_+(x)$ or $u_-(x)$, corresponding to almost all of
the domain being at $1$ or $-1$, respectively, except for a small
boundary layer, since the boundary is kept at $0$ via Dirichlet
boundary conditions.

The natural question is to ask by what mechanism the model switches
from $u_-$ to $u_+$. This is a classical example of nucleation theory,
with applications in phase separation or magnetic field reversal, and
has found particular attention in the small-noise regime and large
deviation theory~\cite{kohn-reznikoff-vanden-eijnden:2005,
  reznikoff-vanden-eijnden:2005, vanden-eijnden-westdickenberg:2008,
  otto-weber-westdickenberg:2014,
  rolland-bouchet-simonnet:2016}. Intuitively, a critical nucleus must
be formed to drive parts of the system over the potential
barrier. After this, front propagation ensures prevailing of the new
phase throughout the whole domain. Since the boundaries are kept at
$0$, nucleating from one of the boundaries is generally easier, so
that the system can flip either via a traveling wave from the left or
from the right. Depending on the size of the domain $L$ (or
equivalently the diffusion constant $\nu$), a nucleus might also form
within the domain.

Since we already start with a 2-dimensional SPDE (space and time), the
corresponding gradient diffusion in pathspace will yield a
3-dimensional SPDE (space, time, and virtual time) for
$\phi(x,t,\tau):[-L,L]\times[0,T]\times[0,\infty]\to \RR$, given by
\begin{equation}
  \partial_\tau \phi = \partial_t^2 \phi - \left(\frac{\delta b}{\delta \phi} - \frac{\delta b^\dagger}{\delta \phi}\right)\partial_t \phi - \frac{\delta b^\dagger}{\delta \phi} b(\phi) -\tfrac12 \eps \frac{\delta}{\delta\phi}\left(\mathrm{Tr} \frac{\delta b}{\delta \phi}\right) + \sqrt{2\eps}\eta(x,t,\tau)\,,
\end{equation}
subject to the boundary conditions
\begin{equation}
  \phi(x,t\!=\!0,\tau) = u_-(x)\,,\quad \phi(x,t\!=\!T,\tau) = u_+(x)\,,\quad \phi(x\!=\!-L,t\tau)=0\,,\quad \phi(x\!=\!L,t\tau)=0\,.
\end{equation}
Here, derivatives are replaced by functional derivatives, and
${}^\dagger$ is the $L^2$-adjoint in the spatial variable.

Note that the $\mathcal O(\eps)$-term, which is the gradient of a
divergence in the finite dimensional case, now includes the
(functional) trace of the operator $\delta b/\delta
\phi$. Unfortunately, this operator is not necessarily trace class,
and the corresponding term might diverge. This is indeed the case for
the Ginzburg-Landau-type cubic nonlinearity
in~(\ref{eq:ginzburg-landau-mcmc}), and is problem well-known and
discussed in the field theory literature~\cite{zinn-justin:2002,
  hochberg-molina-paris-perez-mercader-etal:1999,
  hochberg-molina-paris-perez-mercader-etal:2000}. We follow common
practice~\cite{zinn-justin:2002} here and drop the diverging term, but
remark in passing that this does not rest on any rigorous footing, and
in fact a renormalized term of $\mathcal O(\eps)$ might be more
appropriate.

Concretely, for our choice of $b(\phi) = \nu
\partial_x^2\phi - \phi^3 + \phi$ and given boundary conditions, this
amounts to
\begin{equation}
  \label{eq:ginzburg-landau-mcmc}
  \partial_\tau \phi = \partial_t^2 \phi - (\nu\partial_x^2 -3\phi^2 + 1)(\partial_x^2\phi -\phi^3 + \phi) + \sqrt{2\eps}\eta(x,t,\tau)\,,
\end{equation}
with boundary conditions
\begin{equation}
  \phi(x,t\!=\!0,\tau) = u_-,\quad \phi(x,t\!=\!T,\tau)=u_+,\quad \phi(x\!=\!-L,t,\tau)=0,\quad \phi(x\!=\!L,t,\tau) = 0\,.
\end{equation}
As can be seen by the missing $\partial_t\phi$-term, this system is
actually reversible, but we will add a term breaking reversibility
below.

Integrating~(\ref{eq:ginzburg-landau-mcmc}) corresponds to a Markov
chain Monte Carlo for the Ginzburg-Landau equation, and indeed this
samples from the transition ensemble from $u_-$ to $u_+$. Again,
though, starting the system close to one of transition channels (for
example nucleating from the left boundary as in
figure~\ref{fig:ginzburg-landau-gamma-0}, top left) means that the
sampler will effectively never leave its vicinity. Employing pathspace
metadynamics, instead, results in exploring all possible transition
channels.

As collective variables, we suggest $m=2$ with
\begin{equation}
  f[\phi(x,t)] = (s_1,s_2) = \big(-\int_{-L}^L \phi(x,T/2) \sin(\pi x/2L), \int_{-L}^L \phi(x,T/2) \cos(\pi x/2L)\big)\,,
\end{equation}
taking the $t=T/2$-configuration and projecting it onto $-\sin$ and
$\cos$, respectively. This yields a value for the left-right asymmetry
as $s_1$, and for the center-boundary asymmetry as $s_2$. For example,
nucleating from the left boundary means that around $t=T/2$ most mass is
concentrated in the left half-space, yielding a positive value for
$s_1$ (and roughly 0 for $s_2$).

The converged biasing potential in the CV-plane is depicted in
figure~\ref{fig:ginzburg-landau-UBias} (left) and reveals the
possible transition channels: The landscape clearly has 4 distinct
peaks, corresponding to a nucleation from the left boundary (L), from
the right boundary (R), from both boundaries at the same time (B), and
from the center (C). L and R have identical weight by symmetry, and
are clearly preferred over C and
B. Figure~\ref{fig:ginzburg-landau-gamma-0} shows samples of the four
individual transition scenarios as individual realizations of the
sampler within the four regions, respectively.

The situation becomes more complicated when adding additionally an
advection term to the Ginzburg-Landau equation, namely
\begin{equation}
  \label{eq:ginzburg-landau-shear}
  \partial_t u(x,t) = \nu\partial_x^2 u - u^3 + u - v\partial_x u + \sqrt{2\eps} \eta(x,t)\,,\quad u(-L,t)=u(L,t)=0\,,
\end{equation}
with velocity field $v(x)$ chosen here to be $v(x) =
\gamma\exp(-\frac{1}{1-(x/L)^2})$. Since for $\gamma>0$ this velocity
field is positive within the domain, and only drops to $0$ at the
boundaries, one expects it to favor nucleation from the left boundary:
The rightward advection helps pushing the phase front across the
domain. Nucleation from the right boundary, on the other hand, is
suppressed, since the nucleus must expand against the background
flow. Nucleation in the center, or from both boundaries, will be
distorted by the advection in one direction. Note further that the
advection term is not $L^2$ self-adjoint, and thus the resulting
system is no longer reversible. This is intuitively clear since one
expects the most likely forward transition from $u_-$ to $u_+$ to
nucleate from the left boundary, but backward from $u_+$ to $u_-$ to
nucleate from the left boundary as well, instead of the time-reversed
forward transition.

This intuition is confirmed by the biasing potential, as visible in
figure~\ref{fig:ginzburg-landau-UBias}~(right). Here, the nucleation
from the left boundary (L) is clearly preferred over R for a value of
$\gamma=0.03$. Increasing $\gamma$ further leads to a disappearance of
$R$, while $C$ and $B$ merge with $L$ into the only surviving
transition channel.

For this section, the system parameters are $L=1$, $T=16$, $\nu=4\cdot
10^{-2}$, $\eps=10^{-3}$, $\gamma$ as indicated, and the numerical
parameters are $\Delta \tau = 10^{-2}$, $w=10^{-1}$, $\delta =
10^{-1}$, $\kappa=10^{-1}$, $N_x = 32$, $N_t=32$. The SPDE was
discretized with second order finite-differences in time and Chebyshev
in space, with first order exponential time differencing
(ETD)~\cite{kassam-trefethen:2005} as integrator in virtual time.

\section{Discussion and relation to previous results}
\label{sec:discussion}

We demonstrated a novel algorithm to sample transition trajectories in
complex stochastic systems. The method makes use of the fact that the
transition path ensemble can be sampled efficiently via Langevin MCMC
and accelerated with metadynamics. This makes sure that the algorithm
samples from all possible transition paths even in the presence of
multiple competing transition channels. The flexibility comes at the
price of choosing the right collective coordinates: Without correct
CVs, metadynamics is ineffective or yields misleading
results. Choosing suitable CVs necessitates physical intuition about
the system at hand, and it is not always easy to find a good set of
CVs. The problem of finding CVs for generic sampling problems
is classical, and essential not only for our algorithm here, but for
all rare event algorithms. It can be shown that the optimal collective
variable is the committor function, which is usually not available for any
problem of practical relevance. As mentioned in the introduction, good heuristics
for CVs can come from physical intuition, and machine learning and data science
techniques might offer more systematic insight~\cite{thiede-giannakis-dinner-etal:2019,
  finkel-webber-gerber-etal:2021, lucente-herbert-bouchet:2022,
  lucente-rolland-herbert-etal:2022}.

The presented approach requires storing in memory the whole pathway,
posing an intrinsic limitation to the size of the systems which can be
studied with it. However, because of the increasing speed of modern
computers, we demonstrate that even stochastic partial differential
equations are feasible to be treated without any special effort or
hardware. Since the underlying pathspace Langevin equation is the same
as the gradient decent equation for action minimization (for example
for instanton computation in the Freidlin-Wentzell sense), similar
optimizations can be employed to sample transition paths in even
higher dimensional PDEs, as demonstrated for example for the 3D
Navier-Stokes equation in~\cite{schorlepp-grafke-may-etal:2022}.

Metadynamics has been suggested for use in pathspace
before~\cite{borrero-dellago:2016, mandelli-hirshberg-parrinello:2020,
  bolhuis-swenson:2021}. Generally in those works, the $\mathcal
O(\eps)$-term of the Onsager-Machlup action~(\ref{eq:OM}) is omitted,
even though it has been pointed out in other
contexts~\cite{pinski-stuart:2010, gladrow-keyser-adhikari-etal:2021} 
that the term is crucial for the
correct behavior of transition paths at non-zero temperature. 
Here, we chose to discretize the continuous limiting SPDE, which must
include this term, but conceivably one might alternatively discretize the
original stochastic process first and then derive the bridge sampler fully
discretely. This might allow to sidestep issues of the renormalizability of
the bridge sampling SPDE, in particular for higher dimensional continuous
problems. 

All previous applications of metadynamics to pathspace restrict the
procedure to the setup of equilibrium systems and gradient flows,
where $b(X) = -\nabla V(X)$. As shown in our examples above, the
method presented in this work is generally applicable to a wide range
of situations, including irreversible dynamics without underlying
potential landscape, explicit time dependence of the drift (or noise)
terms, etc. Due to the efficiency of metadynamics, it works even for
high- or infinite-dimensional systems such as SPDEs. Additionally, our
method is readily generalized to situations such as computing
expectations over transition paths, sampling time-periodic paths,
sampling trajectories that visit a sequence of points, sampling
trajectories with a time-integrated constraint, and many other
scenarios encountered in different problem domains.

\paragraph{\bf Acknowledgments}

TG wants to thank Timo Schorlepp for helpful discussions. 
TG acknowledges support from EPSRC projects EP/T011866/1 and EP/V013319/1.

\bibliographystyle{apsrev4-1}
\bibliography{bib}

\end{document}